\documentclass[unnumsec,webpdf,contemporary,large]{oup-authoring-template}%
\usepackage{multirow}
\usepackage{booktabs} 

\usepackage{threeparttable}
\usepackage{algorithm}

\graphicspath{{Fig/}}

\theoremstyle{thmstyleone}%
\theoremstyle{thmstyletwo}%
\theoremstyle{thmstylethree}%
\newtheorem{definition}{Definition}

\begin{document}

\journaltitle{The Computer Journal}
\DOI{DOI HERE}
\copyrightyear{2022}
\pubyear{2019}
\access{Advance Access Publication Date: Day Month Year}
\appnotes{Paper}

\firstpage{1}


\title[Concurrency Bug Detection]{Deep Learning Based Concurrency Bug Detection and Localization}

\author[1]{Zuocheng Feng\ORCID{0009-0003-6695-3533}}
\author[1]{Kaiwen Zhang\ORCID{0000-0003-4573-8968}}
\author[3]{Miaomiao Wang\ORCID{0009-0004-7738-4927}}
\author[1]{Yiming Cheng\ORCID{0009-0007-7021-365X}}
\author[2]{Yuandao Cai\ORCID{0000-0001-6340-1416}}
\author[3]{Xiaofeng Li\ORCID{0009-0002-8805-665X}}
\author[1,$\ast$]{Guanjun Liu\ORCID{0000-0002-7523-4827}}

\authormark{Zuocheng Feng et al.}
\address[1]{\orgdiv{The Key Laboratory of Embedded System and Service Computing of Ministry of Education}, \orgname{Tongji University}, \orgaddress{ \postcode{201804}, \state{Shanghai}, \country{China}}}
\address[2]{\orgdiv{Hong Kong University of Science and Technology}, \orgname{Hong Kong University}, \orgaddress{ \postcode{999077}, \state{Hong Kong}, \country{China}}}
\address[3]{\orgdiv{The Space Optoelectronic Measurement and Perception Lab}, \orgname{Beijing Institute of Control Engineering}, \orgaddress{\street{Street}, \postcode{100086}, \state{Beijing}, \country{China}}}

\corresp[$\ast$]{Corresponding author. \href{email:email-id.com}{liuguanjun@tongji.edu.cn}}

\received{Date}{0}{Year}
\revised{Date}{0}{Year}
\accepted{Date}{0}{Year}



\abstract{Concurrency bugs, caused by improper synchronization of shared resources in multi-threaded or distributed systems, are notoriously hard to detect and thus compromise software reliability and security. The existing deep learning methods face three main limitations. First, there is an absence of large and dedicated datasets of diverse concurrency bugs for them. 
Second, they lack sufficient representation of concurrency semantics. Third, binary classification results fail to provide finer-grained debug information such as precise bug lines. To address these problems, we propose a novel method for effective concurrency bug detection as well as localization. We construct a dedicated concurrency bug dataset to facilitate model training and evaluation. We then integrate a pre-trained model with a heterogeneous graph neural network (GNN), by incorporating a new Concurrency-Aware Code Property
Graph (CCPG) that concisely and effectively characterizes concurrency semantics. To further facilitate debugging, we employ SubgraphX, a GNN-based interpretability method, which explores the graphs to precisely localize concurrency bugs, mapping them to specific lines of source code.
On average, our method demonstrates an improvement of 10\% in accuracy and precision and 26\% in recall compared to state-of-the-art methods across diverse evaluation settings.}


\maketitle

\section{Introduction}\label{sec1}
Concurrency bugs are caused by access to shared resources in a multi-threaded environment without proper synchronization control. 
Concurrent programs are prone to bugs due to the challenges of writing concurrent programs and the non-determinism of concurrent program execution. 
Concurrency issues, such as memory contention, pose significant risks to system integrity, resulting in data corruption, undefined behavior, and potential security breaches. 
Classical static and dynamic analysis methods~\cite{Liu2017,Liu2018,Cai2021,Cai2022,dynamicCai} have been proposed to detect concurrency bugs. 
However, static analysis is typically inaccurate with many false alarms, while dynamic analysis has low recall and code coverage.

Recent studies have leveraged deep learning (DL) to identify potential concurrency bugs in source code~\cite{deeprace,Zhang2021,zhang22,LLMdatarace}.
DL for bug detection (binary classification task to determine whether a bug exists or not) and localization (down to the the specific lines of code) are typically divided into token-based and graph-based methods. Token-based approaches process input data by splitting it into tokens as the basic unit for representation and computation. Graph-based methods use GNNs to model program structures. 

However, current DL methods face three key limitations: 
(i) insufficient representation of concurrency semantics; 
(ii) lacking specialized datasets for concurrency bugs; and (iii) binary outputs that fail to pinpoint fine-grained debug information like exact bug locations.

First, current token-based and graph-based methods cannot model concurrency semantics effectively. 
Token-based methods struggle to capture concurrency semantics and model complex thread interactions due to their reliance on sequential representations, while graph-based methods, despite leveraging structural code information, face challenges in accurately expressing intricate concurrency relationships.

On the one hand, a novel token-based approach named DeepRace~\cite{deeprace}, utilizes a Convolutional Neural Network (CNN) to detect data race in OpenMP and POSIX source code. However, DeepRace only considers the node type of the unfolded Abstract Syntax Tree (AST) in its sequence, which causes a loss of concurrency semantic information and leads to inaccurate detection.
Thapa et al.~\cite{chan2023transformerbasedbugdetectioncode} use Transformer-based methods to detect bugs in sliced and functional code. However, the model fails to capture the structural semantics and syntax of source code.

On the other hand, in contrast to token-based methods, graph-based approaches better capture relationships between nodes in source code. Graphical representations such as AST, Control Flow Graph (CFG), Program Dependency Graph (PDG), and Code Property Graph (CPG)~\cite{cpg14} are common forms of graphcial representations used to analyze program structure. Although graph-based methods~\cite{Zhou19,DeepWukong,ReGVD,VDoTR} have demonstrated significant success in bug detection, they often overlook the inherent concurrency relationships within programs. The expressive power of graph-based representations is limited, making it difficult to model complex concurrency patterns such as thread interactions and hierarchical locking mechanisms. Large-scale program graphs often include noise and redundant information, obscuring critical features and hindering learning. 
As a result, existing graph-based methods also suffer from limited generalizability and scalability. 

Second, current DL approaches for bug detection focus primarily on binary classification. 
As a result, binary classification models in concurrency bug detection hinder further debugging by not being able to provide precise bug locations or detailed insights for developers. Furthermore, the location of a detected bug does not always correspond to the presence of an actual bug, which can result in numerous false positives. 

Third, besides the above limitations, 
there is a significant challenge lies in the scarcity of high-quality concurrency bug datasets. Existing work on concurrency bug detection using DL methods~\cite{deeprace,zhang22,LLMdatarace} is limited and lacks a standardized dataset. Moreover, existing approaches rely solely on data race detection, limiting their effectiveness in capturing broader concurrency issues. The DeepRace dataset lacks location labels; the other datasets address general bugs, and there is currently no specialized dataset for various concurrency bugs.

In this paper, we develop Convul, a novel concurrency bug detection method that integrates the pre-trained model for node embedding and Relational Graph Convolution Network (RGCN)~\cite{RGCN} for bug detection. Our approach combines the semantic understanding of pre-trained models with the relational reasoning power of RGCN, allowing Convul to achieve superior performance in detecting complex bugs with high precision and recall.

First, to address the limitation of existing DL methods in effectively modeling concurrency semantics, we propose Concurrency-Aware Code Property Graph (CCPG), which extends the traditional CPG (a graph representation of the program obtained by merging its AST, CFG and PDG) by incorporating concurrency semantics. We enhance the semantic information of the original CPG by constructing blocking nodes and edges associated with synchronization primitives.

Second, 
we extract a total of 41,317 concurrency bug instances from Diversevul, Big-Vul, SARD, and DeepRace, followed by a manual review to verify that the records accurately correspond to concurrency bugs.
To the best of our knowledge, this is the first dataset specifically designed to detect concurrency-related bugs. Additionally, we identify Common Vulnerabilities and Exposures (CVEs) and CWEs associated with concurrency issues to better understand the characteristics of concurrency bugs and filter out concurrency-related cases from general bugs.

Finally, to address the limitation that binary classification models cannot provide fine-grained information, we propose a novel bug localization model that leverages SubgraphX~\cite{subgraphX} to accurately pinpoint bugs in source code. 
To enhance the algorithm's efficiency, we optimize the objective function for identifying the optimal subgraph and prioritize the search for concurrency-related subgraphs. Our approach effectively captures the intricate graph structural relationship and offers structured and insightful explanations through subgraph analysis.

We compare Convul against eight state-of-the-art (SOTA) DL bug detection
methods and five SOTA DL bug localization methods.

The contributions of this paper are as follows:
\begin{itemize}
\item We create a dataset for concurrency bugs. 
The dataset focuses on precise bug localization, while the data extraction method maps CWEs to concurrency categories, improving detection and analysis.
\item We propose CCPG to address the deficiency in modeling concurrency semantics in existing bug detection methods, which enhances the expressiveness of CPG by incorporating detailed concurrency relationships. CCPG improves the detection of concurrency-related bugs.
\item 
We propose Convul, a novel approach that integrates CodeBERT~\cite{codebert} with RGCN. The integration leverages CodeBERT's semantic understanding and RGCN's structural relationship modeling, enhancing the accuracy and robustness of bug detection.
\item 
We propose a novel bug localization module integrated with Convul, which offers structured location information of concurrency bugs through connected subgraphs and ensures that the predictive performance of the model remains robust and unimpaired.
\end{itemize}

Experiments show that Convul significantly outperforms existing methods in concurrency bug detection and localization, achieving a detection F1-score of 86\% and a localization Intersection over Union (IoU) of 15\%, compared to 73\% and 10\% respectively for the best alternative method.


\section{Method}\label{sec2}
\begin{figure*}[]\centering
\includegraphics[width=\textwidth]{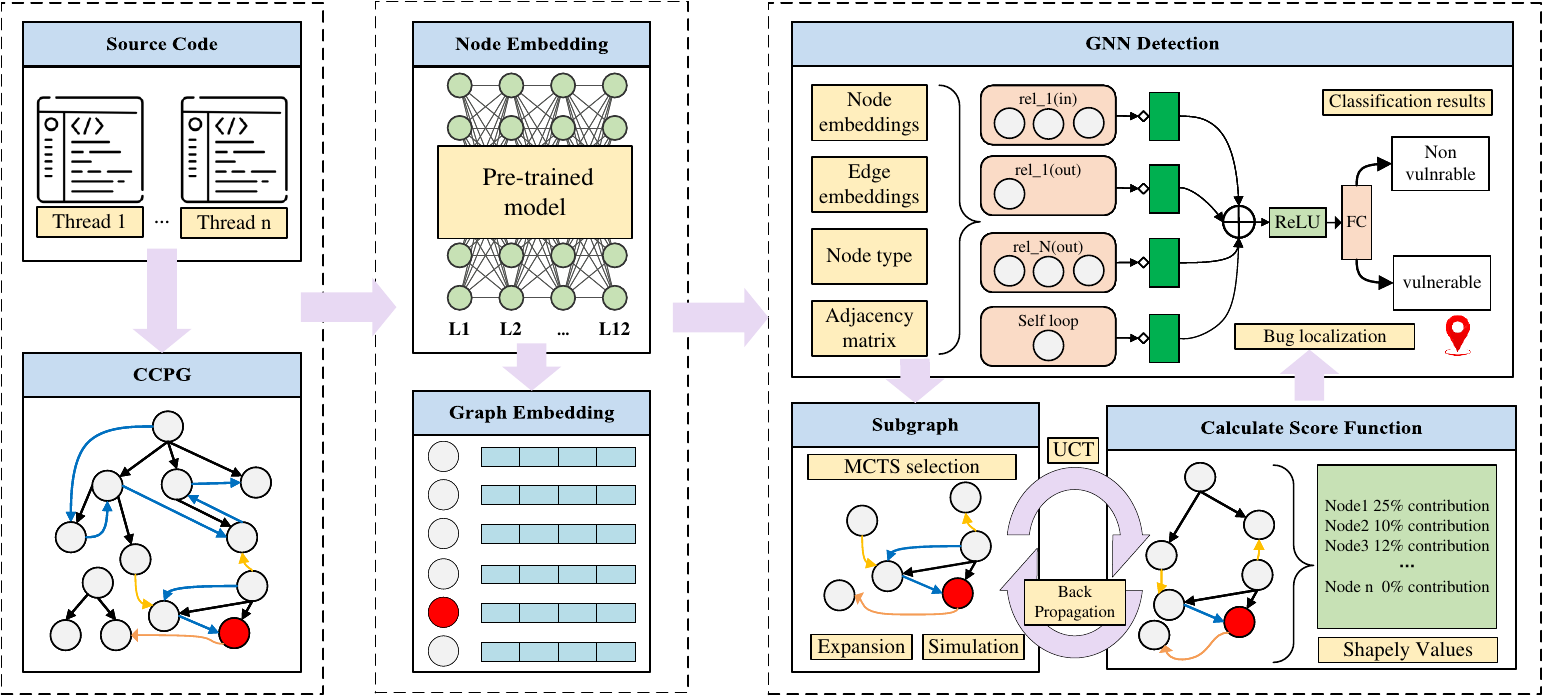}
	\caption{Overview for Concurrency Bug Detection Framework}
	\label{fig:framework}
    
\end{figure*}

Figure \ref{fig:framework}  illustrates the overall framework of our approach. First, we construct CCPGs by extending traditional CPGs to incorporate concurrency-specific attributes. Second, we design a graph embedding method that integrates GNNs with pre-trained models to enhance detection and localization capabilities. Finally, we integrate the interpretable method SubgraphX into the GNN, leveraging Monte Carlo Tree Search (MCTS) to identify subgraphs containing nodes with bugs for precise bug localization.


\subsection{Dataset}\label{subsec1}
In this subsection, the method for extracting datasets for concurrency bugs will be introduced.

\begin{table}[!t]
\caption{Summary of bug Datasets}
\label{tab:datasets}
\centering
\begin{threeparttable} 
\resizebox{1.0\linewidth}{!}{  
\begin{tabular}{ccccr}
\toprule
Dataset    &Granularity& CWE/CVE $^{\mathrm{1}}$                 & Loc Info$^{\mathrm{2}}$ & Total  \\ \midrule
VulDeePecker~\cite{vuldeepecker}  &  Slice& \checkmark & $\times$ & 61,638   \\
SeVC~\cite{SySeVR}    &Slice& \checkmark & $\times$ & 420,627   \\
FFMPeg+Qemu~\cite{Zhou19}     & Function& $\times$    &$\times$     &27,318                       \\
\textbf{DiverseVul}~\cite{DiverseVul} &Function & \checkmark & $ \times $ & 330,492  \\
\textbf{BigVul}~\cite{BigVul}  & Function& \checkmark & \checkmark & 217,007 \\
Reveal~\cite{Reveal}  &Function& $ \times $\ & $ \times $\ & 22,734  \\
D2A~\cite{D2A} &Trace& $ \times $\ & \checkmark & 1,314,276  \\
\textbf{SARD}~\cite{SARD}  &  Function& \checkmark & \checkmark & 426,654  \\
\textbf{Deeprace}~\cite{deeprace}    &File& $\times$                           &  $\times$       &   16,085                     \\ \botrule
\end{tabular}
}
    \begin{tablenotes}    
        \footnotesize               
        \item[1] Whether it contains the corresponding CWE or CVE tag.         
        \item[2] Whether it contains information about the location of bugs.     
    \end{tablenotes}   
\end{threeparttable}

\end{table}

CVE-2024-50066~\cite{CVE-2024}, contains detailed information about a bug linked to a unique CVE ID.
The initial screening process identifies concurrency-related CVE records based on the presence of specific keywords within their descriptions. The keywords include terms such as lock, deadlock, mutex, spinlock, semaphore, synchronization, thread, multi-thread, multiple-threads, concurrency, concurrent, parallel, atomicity, order violation, race, 
among others. 
The process yields a CVE list spanning from 1999 to 2024, consisting of 1,345 CVE records. For CWE records, the analysis focuses on 
CWE-1401~\cite{CWE1401}. Currently, there are 37 records classified under CWE-1401.

\begin{figure}[!h]\centering
\includegraphics[width=0.48\textwidth]{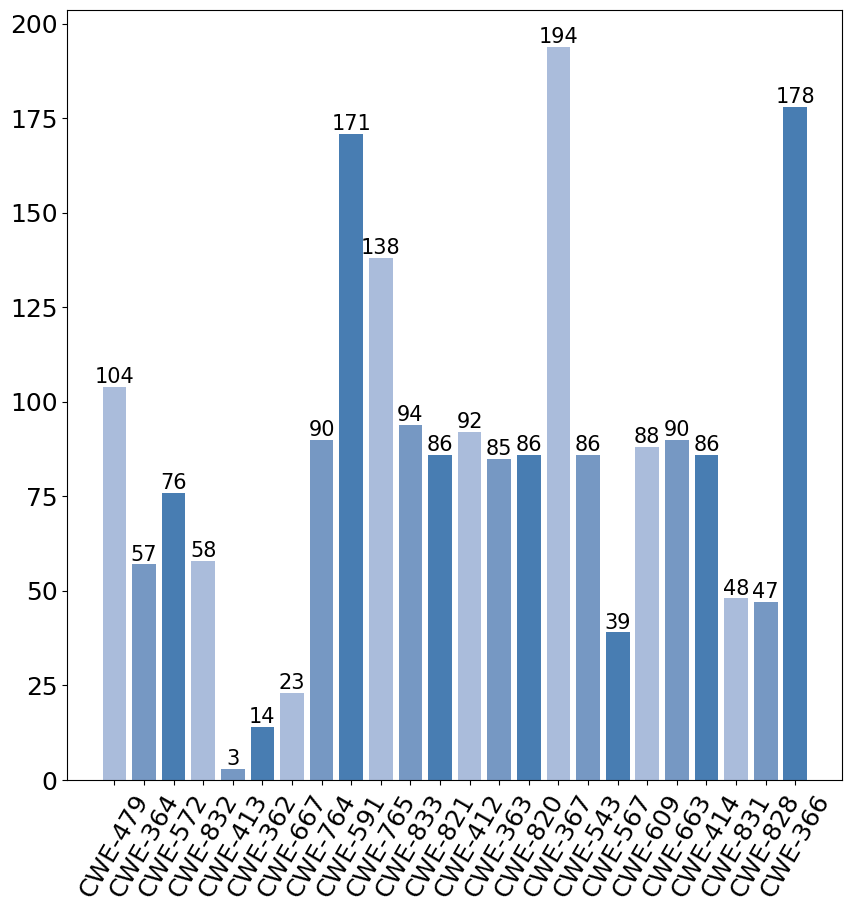}
	\caption{Distribution of Concurrent Vulnerabilities by CWE in SARD}
	\label{fig:SARD}
\end{figure}

To facilitate the filtering of concurrency bugs, we ensure that dataset entries include CWE or CVE labels. We thoroughly examine existing open-source bug detection datasets, as summarized in Table \ref{tab:datasets}.
These samples serve as the foundational elements driving our experimental research. The concurrency bugs extracted from Diversevul, Bigvul, SARD, and DeepRace can be consolidated to construct a novel concurrency bug dataset. Additionally, concurrency bugs extracted from BigVul and SARD can be used in bug localization. 
The distribution of the SARD dataset across concurrent CWEs is illustrated in Figure \ref{fig:SARD}. The analysis of the data distribution presented in the figure indicates that the incidence of concurrency bugs is predominantly concentrated in the category of data races.

\subsection{Code Representation with CCPG}
\begin{algorithm}[!t]
\caption{Simplified Algorithm for Creating CCPG}
\label{alg:Alg1}
\renewcommand{\algorithmicrequire}{\textbf{Input:}}
\renewcommand{\algorithmicensure}{\textbf{Output:}}

\begin{algorithmic}[1]
\Require Code Property Graph $CPG=(G, E)$  
\Ensure Concurrency-Aware Code Property Graph $CCPG$

\State Define synchronization primitives SP = \{lock, unlock\}  
\State Define thread control functions TCF = \{create, join\} 
\State Initialize function node map $FN = f:(start, last)$  

\For{each function graph $g \in CPG$}
    \State Add start and end nodes $(g_s, g_l)$ of $g$ to $FN$  
    \State Add $g$ to $CCPG$  
\EndFor

\For{each $g \in CCPG$ and $v \in g$}
    \If{$v \in TCF.create$}
        \State Add edge $(v \to FN[g].s)$ to $E_{sync}$  
    \ElsIf{$v \in TCF.join$}
        \State Add edge $(FN[g].l \to v)$ to $E_{sync}$ 
        \State Add $v$ to $V_{block}$
    \ElsIf{$v \in SP.lock$}
        \State Track $v$ as a lock/unlock node to $V_{block}$
        \State Add edge $(v_1\in V_{lock} \to v_2\in V_{unlock})$ to $E_{sync}$
    \EndIf
\EndFor
\State \Return $CCPG$
\end{algorithmic}
\end{algorithm}

A CPG represents a program as a unified graph obtained by merging its AST, CFG, and PDG at statement and predicate nodes. The resulting structure is a property graph, a model used in graph databases like Joern \cite{Joern}, where data is stored in nodes and edges as key-value pairs. Consequently, code property graphs can be stored in graph databases and queried using graph query languages, facilitating efficient analysis and manipulation.

\begin{definition}
A \textit{Concurrency-Aware Code Property Graph} is a 2-tuple $G=(V,E)$, where: $V=\{V_{block} \cup V_{unblock}\}$ is a finite set of vertices, $V_{block}$ is the set of vertices in the critical regions and $V_{unblock}$ is the set of vertices corresponding to operations that release synchronization primitives. $E=\{E_{cpg} \cup E_{sync}\}$ is a finite set of edges, $E_{sync}$ represents the set of edges associated with synchronization primitives and $E_{cpg}$ includes all edges from the CPG. $V \cap E = \emptyset$.
\end{definition}

We extend the CPG to a CCPG by incorporating concurrency edges and blocking nodes, as detailed in Algorithm \ref{alg:Alg1}.
First, all methods in the CPG are traversed to identify their entry and exit nodes, integrating complete method structures into the CCPG. Next, the call graph is analyzed to detect calls to POSIX functions such as $pthread\_create$ and $pthread\_join$. The convergence edges are established from the caller to the specific thread function extracted from $pthread\_create$, while the edges are directly connected to $pthread\_join$ due to the lack of function-specific information and pointer analysis capabilities in Joern. Finally, within each method's CFG, critical sections are identified by traversing paths between $pthread\_mutex\_lock$ and $pthread\_mutex\_unlock$. Nodes along these paths are marked as blocking nodes, and their corresponding edges are labeled as concurrency edges to reflect synchronization behavior. CCPG enhanced by additional nodes and edges offers a comprehensive representation of concurrency, facilitating more effective analysis of multi-threaded programs and detection of potential concurrency issues.

\begin{figure*}[!htb]\centering
	\includegraphics[width=\textwidth]{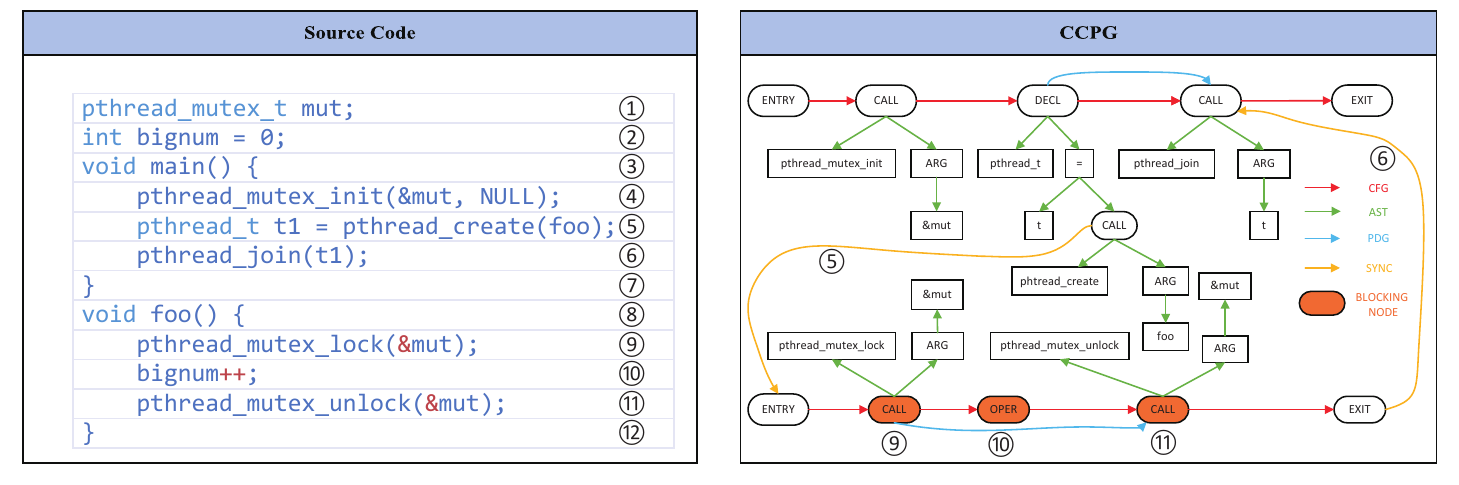}
	\caption{An Example of CCPG}
	\label{fig:cpg_extend}
\end{figure*}

An example of a transformation is shown in Figure \ref{fig:cpg_extend}, with part of the code on the left that creates a thread in the main function to manipulate global variables $bignum$, which is performed in the critical section.


\subsection{ Node and Graph Embedding}
\begin{algorithm}[!t]\small
\caption{Concurrency bug Detection Using CCPG}\label{alg:RGCN}
\begin{algorithmic}[1]
\renewcommand{\algorithmicrequire}{\textbf{Input:}}
\renewcommand{\algorithmicensure}{\textbf{Output:}}
\Require CCPG $G = (V, E)$. 
\Ensure Graph Representation $\mathcal{R}_G$ at layer $l$.

\State \textbf{Input Graph Preprocessing:}
\State Construct the CCPG with Algorithm \ref{alg:Alg1}.

\State \(\rhd\) \textbf{Node Feature Embedding:}
\ForAll{$v \in V$}
    \State $h_v^{(0)} \gets CodeBERT(v)$ 
\EndFor

\State \(\rhd\) \textbf{Graph Representation Learning:}
\For{$l = 1, 2, \dots, L$}
    \ForAll{$v \in V$}
        \State Update node embeddings following Eq.\eqref{eq:RGCNupdate}
    \EndFor
\EndFor

\State \(\rhd\) \textbf{Readout and Pooling:}
\State $\mathcal{R}_G \gets \text{readout}(\{h_v^{(l)} : v \in V\})$ 
\State \Return $\mathcal{R}_G$.

\end{algorithmic}
\end{algorithm}
In this subsection, we initialize node embeddings with CodeBERT which effectively captures the semantic information of the code.
After constrcuting CCPG, we transform the code tokens of each
CCPG's node into a vector.
Traditional embedding methods, such as word2vec~\cite{word2vec}, produce static embeddings where a word or node retains the same vector representation across all contexts, limiting the capture of nuanced semantic information. In contrast, using the pre-trained model CodeBERT to generate initial node embeddings enables the representation of richer semantic and syntactic information. In addition, CodeBERT can be fine-tuned to adapt to domain-specific data, enhancing its versatility. In a CCPG, there are nodes \( N = \{ n_1, n_2, \ldots, n_n \} \) that correspond to the sequence of statements \( S = \{ s_1, s_2, \ldots, s_n \} \). CodeBERT employs Byte Pair Encoding (BPE) for subword segmentation, effectively addressing the diverse range of programming language identifiers, such as variable and function names. BPE also captures similarities and distinctions among programming languages, making it well-suited for code representation tasks. The task for CodeBERT to generate node embeddings can be formalized as:
\begin{eqnarray}
    \mathcal{V}_i\mathit{} = CodeBERT(BPE\mathit{}(s_{i}))\mathit{}
\end{eqnarray}
where \(\mathcal{V}_i\) is final embedding of node \(n_i\).

To represent edges as a two-dimensional matrix for input to the GNN, we employ the Coordinate Format (COO), a storage format optimized for sparse matrices. The COO format is typically represented as a triple \((row, column, data)\), where \textit{row} denotes the row index of a non-zero element, \textit{colum} specifies its column index, and 
\textit{data} stores the value of the element. Nodes are assigned numerical indices using depth-first traversal (DFS).

\subsection{RGCN Based Feature Learning}
In this work, we adopt RGCN to handle the graph representations of concurrent program. we conclude the computation steps of RGCN with CCPG in Algorithm \ref{alg:RGCN}.
In concurrency bug detection, the complexity and diversity of code structures present significant challenges for traditional static analysis methods, often hindering their ability to comprehensively identify potential security risks. RGCN, a heterogeneous GNN, is well-suited for detection task due to its capability to effectively process multiple types of graph data. 

Considering the multiple relationships of heterogeneous graphs, the forward propagation of RGCN is formulated as:
\begin{eqnarray}
{h}_i^{(l+1)} = \sigma \left( {W}_0^{(l)} {h}_i^{(l)} + \sum_{r \in \mathcal{R}} \sum_{j \in \mathcal{N}_i^r} \frac{1}{z_{i,r}} {W}_r^{(l)} {h}_j^{(l)} \right)
\label{eq:RGCNupdate}
\end{eqnarray}
where 
\( \mathcal{R} \) is the set of all possible relation types and \( \mathcal{N}_i^r \) denotes the set of neighbor nodes of node \( i \) under relation \( r \). \( z_{i,r} \) is a normalization constant, \(W\) is the trainable weight matrix and \( {h}_j^{(l)} \) is the embedding of neighboring node \( j \) at layer \( l \).

 For a L-layer GNN, representation \(\mathcal{R}\) of graph \(\mathbf{G_k}\)  can be computed using readout function (e.g., mean pooling) as:
\begin{eqnarray}
\mathcal{R}_k(\mathbf{G_k})=\sigma\left(\frac{1}{N}\sum_{i=1}^N h^{l+1}_i\right)
\end{eqnarray}
Finally, we use a Multi-Layer Perception (MLP) followed by a softmax function to predict the probability $P$ of  a conccurrency bug  as follow:
\begin{eqnarray}
P(\mathbf{G_k})=\mathrm{Softmax}(\mathrm{MLP}(\mathcal{R}_k)) 
\end{eqnarray}

\section{Concurrency Bug Localization with SubgraphX}
The graph-based interpretation model is a novel approach for locating concurrency bugs in source code. We use SubgraphX to build our bug localization model, aiming to identify important subgraphs to interpret the GNN model. The detail is shown in Algorithm \ref{alg:subgraphx}. The subgraph illustrates nodes that we believe may have concurrency bugs.

Compared to GNNExplainer \cite{GNNExplainer}, which identifies discrete edges, SubgraphX provides more human-interpretable explanations by generating connected subgraphs. While SubgraphX does not necessarily reduce theoretical computational complexity, it improves practical efficiency in identifying important subgraphs by employing a MCTS strategy, which enables more scalable and interpretable explanations, especially in large-scale graph scenarios. We present a formal explanation of how SubgraphX works.
\subsection{MCTS}
\begin{algorithm}[!t]\small
\caption{Graph-Based Concurrency Bug Localization with MCTS}
\label{alg:subgraphx}
\begin{algorithmic}[1]
\renewcommand{\algorithmicrequire}{\textbf{Input:}}
\renewcommand{\algorithmicensure}{\textbf{Output:}}
\Require Graph $\mathcal{G} = (V, E)$, GNN model $\mathcal{M}$, max iterations $B$, parameters $(\lambda,\alpha,\beta)$, sampling steps $T$.
\Ensure Subgraph $\mathcal{G}_{max}$ indicative of concurrency bug.
\State Initialize $\mathcal{S}_{max} \gets 0$, $\mathcal{T} \gets \text{empty tree}$, root $\mathcal{N}_0$.
\State Initialize $Q(s,a) \gets 0$, $N(s,a) \gets 0$.

    \For{$t = 1$ to $B$}
        \State $\mathcal{N}_{cur} \gets \mathcal{N}_0$, $path \gets [\mathcal{N}_0]$.
        \While{$\mathcal{N}_{cur}$ is not a leaf and $|h(\mathcal{N}_{cur})| \leq SN_{max}$}
            \For{pruning actions}
                \State $\mathcal{N}_{cur} \gets \text{child}(\mathcal{N}_{cur}, a)$.
                \State $S \gets \text{CalculateShapley}(\mathcal{G}, h(\mathcal{N}_{cur}), \mathcal{M}, T)$.
            \EndFor
            \State select $\mathcal{N}_{next}$ following \eqref{eq:argmax}.
            \State $\mathcal{N}_{cur} \gets \mathcal{N}_{next}$, $path \gets path \cup \mathcal{N}_{next}$.
        \EndWhile
        \If{$S > S_{max}$}
            \State $S_{max} \gets S$, $\mathcal{G}_{max} \gets \mathcal{G}_{cur}$.
        \EndIf
        \For{each $(s, a) \in path$}
            \State Backpropagation following \eqref{eq:update1}\eqref{eq:update2}.
        \EndFor
    \EndFor
    
\State\Return $\mathcal{G}_{max}$.
\end{algorithmic}
\end{algorithm}

SubgraphX employs MCTS as efficient search algorithm. SubgraphX constructs a search tree. The root node is associated with the input graph, while each other node corresponds to a connected subgraph. We define a node in a correspond search
tree as \(\mathcal{N}_i\) while \(\mathcal{N}_0\) denotes the root node.
MCTS can be broadly categorized into four steps: Selection, Expansion, Simulation, and Backpropagation.

For a series of subgraphs \(\left\{\mathcal{G}_{1}, \cdots, \mathcal{G}_{i}, \cdots, \mathcal{G}_{n}\right\}\) of the CCPG \(G\),
our goal is to select a optimal subgraph \(\mathcal{G}^{*}_{s}\) which can be defined as:
\begin{eqnarray}
\mathcal{G}^{*}_{s}=\underset{\left|\mathcal{G}_{s}\right| \leq N_{\min},{G}_{s}\in G}{\operatorname{argmax}} \operatorname{\textit{S}}\left(\mathcal{M}(\cdot), \mathcal{G}, \mathcal{G}_{s}\right)
\end{eqnarray}
where \(S(,,)\) represents a scoring function used to assess the significance of a subgraph based on the trained GNNs and the provided input graph. \({N}_{min}\) is an upper limit on subgraph size to ensure that the resulting explanations remain concise and avoid generating overly complex explanations. \(\mathcal{M}(\cdot)\) denotes the classification prediction function of the GNN for subgraph \(\mathcal{G}\) which can be formulated as \(\mathcal{M}:\mathcal{G} \mapsto \mathcal{Y}\) where \(\mathcal{Y}\) denotes the space of sample labels.

Since we aim to concentrate more on concurrency-related subgraphs, we establish a parameter \( \theta  \) that denotes the number of edges \(N_{ce}\) with synchronization primitives  and the number of blocking nodes \(N_{bn}\) within the graph. The parameter \( \theta  \) can be defined as:
\begin{eqnarray}
    \theta =\alpha N_{ce} +\beta N_{bn}
\end{eqnarray}
where  
\(\alpha\) and  \(\beta\) are hyperparameters.
At the selection stage, starting from the root node, we recursively select child nodes until we reach a leaf node.
We assume that the subgraph is derived from \(\mathcal{G}_{i}\) through the pruning operation \(a_j\). For  \(\text{pair}\,({s}_i,a_j)\) of state \({s}_i\) and pruning action \(a_j\), the action selection criterion for a state \({s}_i\) with node \(n_i\) and subgraph \(N_i\) is defined as:


    \begin{eqnarray}
    a^{*}=\underset{a_{j}}{\operatorname{argmax}}\ 
 Q\left({s}_{i}, a_{j}\right)+(\theta +1)U\left({s}_{i}, a_{j}\right)
\label{eq:argmax}
\end{eqnarray}

\begin{eqnarray}
Q\left({s}_{i}, a_{j}\right)=W\left(s_i,a_j\right)/C\left(s_i,a_j\right)
\end{eqnarray}

\begin{eqnarray}
    U\left({s}_{i}, a_{j}\right)= \lambda S\left(\mathcal{M}(\cdot), \mathcal{G}, \mathcal{G}_{j}\right)\frac{\sqrt{\sum_kC( s_i,a_k)}}{1+C(s_i,a_j)}\quad
\end{eqnarray}
where \(W(s_{i},a_{j})\) is the total reward for all \((s_i,a_j)\) visits. \(C(s_i,a_j)\) represents the count of selections for pruning action \(a_j\) at node \(\mathcal{N}_i\). \(\lambda\) is a hyperparameter for balancing trade-offs between current and unexplored nodes.

In the expansion phase, if the selected leaf node \(\mathcal{N}_l\) has not been visited previously, it is expanded.
A subgraph corresponding to \(\mathcal{N}_l \) is obtained through a search and sampling process.
In the simulation phase, for each node in expand nodes, we pruning it and get the remaining subgraph. We calculate score function \(S\) for each subgraph.
In the back propagation phase, we update the cumulative rewards and the number of visits to the node as:
\begin{eqnarray}
    C\left({s}_i,a_j\right) = C\left({s}_i,a_j\right) + 1
\label{eq:update1}
\end{eqnarray}
\begin{eqnarray}
W\left({s}_i,a_j\right)=W\left({s}_i,a_j\right)+ S\left(\mathcal{M}({G}_{l}), \mathcal{G}, \mathcal{G}_{l}\right)
\label{eq:update2}
\end{eqnarray}

\subsection{Compute Score Function with Shapley Value}
Shapley value is a concept from cooperative game theory that assigns a fair contribution value to each player (or feature) based on their marginal contribution to all possible coalitions, providing a method for distributing rewards or costs among participants.
Shapley values enhance the interpretability of model predictions by equalizing the contributions of individual subgraphs to the model's outputs.
In the \( L \)-layer GNN, the subgraph \(\mathcal{G}_{i}\) with nodes \(\{v_1,v_2,\cdots,v_k\}\) will interact with its neighbors for \( L \)-hops. Let \( r \) represent the number of nodes within the \( L \)-hop neighborhood of the subgraph \( G \).
The set of players is defined as
\(P = \{\mathcal{G}_i,v_{k+1},\cdots,v_r\}\).
SubgraphX uses Shapley value to calculate score function. Shapley values were used to quantify the effect of subgraphs on the model's predictions. Here Shapley value is defined as:
\begin{small}
    \begin{eqnarray}
    \varphi(\mathcal{G}_i)=  \sum_{S \subseteq p \backslash\{\mathcal{G}_i\}}\binom{|P|-1}{|S|}^{-1}(\mathcal{M}(S \cup\{\mathcal{G}_i\})-\mathcal{M}(S))
\end{eqnarray}
\end{small}

where S is the possible federated set of participants (excluding \(\mathcal{G}_i\)). \(\binom{n}{k}\) denotes the binomial coefficient.

Furthermore, the Shapley value \(\varphi(\mathcal{G}_i)\) is calculated by considering the neighboring nodes in L-hop, and the Monte-Carlo sampling is used to calculate \(\varphi(\mathcal{G}_i)\). In the sampling step denoted as \(i\), we select a coalition set \( S_i \) from the player set of and calculate its edge contribution as \(\mathcal{M}(S_i \cup\{\mathcal{G}_i\})-\mathcal{M}(S_i)\). We average the results of total \(T\) sampling steps. The final Shapley value can be obtained from the equation as:
\begin{small}
    \begin{eqnarray}
    \varphi(\mathcal{G}_i)=\frac{1}{T}\sum_{t=1}^T(\mathcal{M}\left(S_i\cup\{\mathcal{G}_i\}\right)-\mathcal{M}(S_i))
\end{eqnarray}
\end{small}

\section{Experiments}

Our experimental setup was conducted on a system running Ubuntu 23.10, equipped with a 12th Gen Intel(R) Core(TM) i9-12900K CPU and a GeForce RTX 3090 Ti GPU.
In this section, we present the experimental results and address four key research questions:
\begin{itemize}
    \item \textbf{RQ1} How does Convul compare to existing concurrency bug detection and other general bug detection models?
    \item \textbf{RQ2} What is the impact of different node embedding models and GNNs on the experimental results?
    \item \textbf{RQ3} Does CCPG improve the effectiveness of concurrency bug detection?
    \item \textbf{RQ4} How does our model compare to other bug localization models?
    
\end{itemize}

\subsection{Evalution Metrics}
\label{subsec:evalution metrics}
For detection metrics, we follow the evaluation metrics from
prior bug detection research~\cite{metrics} using \textit{Accuracy} (Acc), \textit{Precision} (Pre), \textit{Recall} (Re) and \textit{F1-score} (F1).
For localization metrics, we use 
\textit{IoU} to measure the extent of nodes overlap between the subgraph \(\mathcal{G}_{pred}\) predicted by the model and the actual set of vulnerable nodes \(\mathcal{G}_{true}\).
The calculation of \(IoU\) can be expressed as follows:
\begin{equation}
\mathrm{\textit{IoU}=\frac1n\sum_{i=1}^n\frac{|\mathcal{G}_{pred}\cap \mathcal{G}_{true}|}{|\mathcal{G}_{pred}\cup \mathcal{G}_{true}|}}
\end{equation}
where \(n\) is the number of graph data samples. Moreover, a subgraph \(\mathcal{G}_{i}\)  is considered accurately localized if it contains at least one bug node, indicating a correct interpretation.
The accuracy is defined as the ratio of accurately localized subgraphs to the total number of subgraphs.

\subsection{Comparison with Detection Baselines (RQ1)}
We compare Convul with existing token-based and graph-based methods for bug detection. The experimental results are shown in Table \ref{tab:model_performance_resized2}. 
On the file-level DeepRace's dataset, Convul demonstrates superior performance, improving accuracy, recall, and F1-score by 17\%, 9\%, and 12\%, respectively, compared to the SOTA graph-based approach while Convul also performs superior to the token-based baseline methods.
For baselines, we select a diverse set of models, including five token-based methods (SySeVR~\cite{SySeVR}, Linevul~\cite{Linevul}, CodeBERT, GraphCodeBERT~\cite{graphcodebert} and DeepRace) and four graph-based methods (i.e., DeepWukong~\cite{DeepWukong}, Devign~\cite{Zhou19}, ReGVD~\cite{ReGVD} and Reveal). The parameter settings for the CodeBERT and GraphCodeBERT are based on this work~\cite{chan2023transformerbasedbugdetectioncode}.

\begin{table}[!t]
\caption{Comparison with Detection Baselines}
\label{tab:model_performance_resized2}
\centering
\resizebox{\linewidth}{!}{  
\begin{tabular}{llllll}
\toprule
    \textbf{Dataset} &  \textbf{Model}& \textbf{Acc} & \textbf{Pre} &\textbf{Re}  &\textbf{F1}    \\ \midrule
\multirow{8}{*}{DeepRace} 
                    &DeepRace &  70.38 &69.68  &76.97  & 73.14   \\
                    & SySeVR & 60.03 & 58.25 & 62.89 & 60.48  \\
                  & Devign &52.47  &54.00  &51.92  &52.94  \\
                         & ReGVD &58.62  &57.93  & 58.92 & 58.42 \\
                         & Reveal & 53.92& 48.68&59.84& 53.69\\
                         & Linevul &64.33 & 57.92 & 92.75 & 71.30\\
                         & CodeBERT & 66.11 & 62.54 & 81.60 & 70.81  \\
                         & GraphCodeBERT & 67.35  & \textbf{84.89} & 39.67 & 54.07   \\
                         &Convul w/o CCPG& 61.83&70.03&77.34& 73.50\\
                         & \textbf{Convul w/ CCPG} & \textbf{75.68} & 78.32 & \textbf{94.60} &  \textbf{85.69}\\
                  \botrule
\end{tabular}
}
\end{table}
For token-based baselines, SyseVR extracts bug-related syntax characteristics through program slicing.
However, the slicing-based approaches primarily focus on sensitive APIs, arrays, pointers, and expressions, with no slicing capabilities for concurrency bugs involving shared variables. Transformer-based methods achieved over 90\% accuracy on merged function-level datasets as showed in Table \ref{tab:Transformer-based model}, while they perform poorly on the standard C/C++ concurrency code dataset DeepRace. 
The Transformer-based seq2seq model exhibits limitations in concurrency bug detection due to its inability to effectively capture complex dependencies and context-specific interactions between concurrent execution paths, leading to suboptimal performance in identifying vulnerabilities in parallel or multi-threaded code structures.

For graph-based baselines, models like Devign and ReGVD are limited to function-level inputs and can't accommodate file-level data. For such models, we use main entry functions as inputs. In contrast, Convul supports input at all levels of granularity, including slice, function, and file level.

\begin{table}[!t]
\caption{Transformer-based model's performance}
\label{tab:Transformer-based model}
\centering
\resizebox{\linewidth}{!}{  
\begin{threeparttable}
\begin{tabular}{llllll}
\toprule
\textbf{Dataset} & \textbf{Model} & \textbf{Acc} & \textbf{Pre} & \textbf{Re} & \textbf{F1} \\ 
\midrule

\multirow{3}{*}{merged dataset$^{\mathrm{1}}$}
& Linevul & 95.25 & 90.65 & 75.67 & 82.48 \\
& CodeBERT & 95.32 & 92.31 & 75.00 & 80.35 \\
& Graphcodebert & 93.03 & 90.62 & 78.56 & 82.76 \\

\bottomrule
\end{tabular}
\begin{tablenotes}
    \footnotesize
    \item[1] merged from Diversevul, Big-Vul and SARD
\end{tablenotes}   
\end{threeparttable}
}
\end{table}

\subsection{Different GNNs and Embedding Models Combination (RQ2)}
To verify the impact of different GNNs on the
bug detection task, we conduct comparative experiments on three different GNNs, include Graph Convolutional Network (GCN)~\cite{GCN}, Graph Attention Network (GAT)~\cite{GAT}
and RGCN. And we compare the detection results on four different node embedding models, include doc2vec~\cite{doc2vec}, word2vec, CodeBERT and GraphCodeBERT.
Both doc2vec and word2vec are configured with a vector dimension of 100, while the two BERT-based models utilize an embedding dimension of 768. The results are shown in Table \ref{tab:d_embedding_model_performance}. Among the embedding models, CodeBERT achieved the highest precision at 88.12\%. It is evident that pre-trained models are more suitable for code embedding than the traditional word2vec and doc2vec models.
Then we compare our model's performance using different GNNs, as shown in Table \ref{tab:gnn_code_performance1}. RGCN achieves an accuracy of 84.90\%, outperforming both GCN and GAT. RGCN effectively captures the features of concurrency heterogeneous relationship by utilizing weight matrices for various relation types, enabling the independent modeling of each relation.

\begin{table}[!t]
\caption{Performance with different embedding models}
\label{tab:d_embedding_model_performance}
\small
\centering
\resizebox{\linewidth}{!}{  
\begin{tabular}{c l c c c c}
\toprule
\textbf{GNN} & \textbf{Embedding Model} & \textbf{Acc} & \textbf{Pre} & \textbf{Re} & \textbf{F1} \\
\midrule
\multirow{4}{*}{RGCN} 
                       & doc2vec & 76.28 & 74.37 & 80.56 & 77.34 \\ 
                       & word2vec & 80.08 & 77.29 & 76.81 & 77.04 \\
                       & CodeBERT & \textbf{84.90} & \textbf{88.12} & 80.31 & \textbf{84.03} \\
                       & GraphCodeBERT & 82.18 & 81.08 & \textbf{86.96}& 89.91 \\
\botrule
\end{tabular}
} 

\vspace{-0.1cm}
\end{table}

\begin{table}[!t]
\caption{Performance with different GNNs}
\label{tab:gnn_code_performance1}
\small
\centering
\resizebox{\linewidth}{!}{  
\begin{tabular}{c l c c c c}
\toprule
\textbf{Embedding Model} & \textbf{GNN} & \textbf{Acc} & \textbf{Pre} & \textbf{Re} & \textbf{F1} \\
\midrule
\multirow{3}{*}{CodeBERT} 
                       & GCN & 73.42 & 75.65 & 74.59 & 75.11 \\ 
                       & GAT & 79.68 & 76.52 & 79.33 & 77.89 \\
                       & RGCN & \textbf{84.90} & \textbf{88.12} & \textbf{80.31}& \textbf{84.03} \\
\botrule
\end{tabular}
} 

\vspace{-0.3cm}
\end{table}
\subsection{Impact
of CCPG (RQ3)}
We conduct an ablation study to evaluate the impact of incorporating CCPG on the performance of the CodeBERT and GNN framework for concurrency bug detection on DeepRace POSIX dataset based on Table \ref{tab:model_performance_resized2}. Specifically, we compared the results obtained with and without the inclusion of CCPG in the graph-based representation.  The results show an improvement of at least 8.29\% in Acc, Pre, Re and F1 when CCPG is applied. 
Constructing concurrency relationships on the CPG to form a CCPG is effective for concurrency bug detection, as this approach captures the complex interdependencies and execution order of different threads, enabling  more accurate identification of potential race conditions and synchronization issues.
\begin{table}[!t]
\caption{Convul's Performance with Localization Baselines}
\label{tab:localization}
\tiny
\centering
\resizebox{\linewidth}{!}{
\begin{tabular}{l c c c c}

\toprule
\multirow{2}{*}{\textbf{Model}} & \multicolumn{2}{c}{\textbf{Big-Vul}} & \multicolumn{2}{c}{\textbf{SARD}}
\\ 
& \textbf{Acc}   & \textbf{\textit{IoU}}   &  \textbf{Acc}   &  \textbf{\textit{IoU}}  \\ 
\midrule
Linevul             &  81.32         &  9.60          &  67.58          &  7.42          \\
Devign              &  79.45         &  6.89          &  45.97          &  3.85          \\
VulDeeLocator       & \textbackslash & \textbackslash &  70.07          &  5.91          \\
VulChecker          &  85.68         &  8.80          &  71.58          & 6.50           \\
IVDetect            &  87.90         &  9.91          &  76.12          & 7.62           \\
\textbf{Convul}        & \textbf{88.30} & \textbf{14.50} & \textbf{79.68 } & \textbf{12.57} \\ 
\botrule
\end{tabular}
}

\vspace{0.1cm}
\end{table}

\begin{figure}[!t]\centering
    \includegraphics[width=0.5\textwidth]{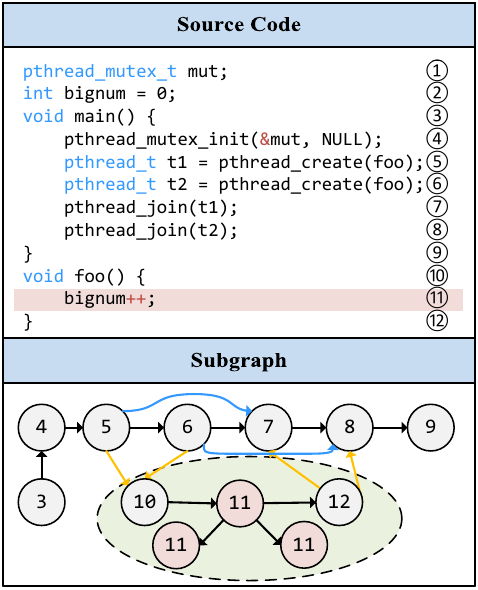}
	\caption{Detection and Localization Motivating Example}
	\label{fig:location motivation}
\end{figure} 
\subsection{Comparison with Localization Baselines (RQ4)}
We conduct comparative experiments using various bug localization models. The experimental results are shown in Table \ref{tab:localization}. 
Our detection experiments incorporate two baseline models: LineVul and Devign. Both models support bug localization, enabling a direct comparison of their performance with other SOTA models. We compare our model with other up-to-date baselines as follows:
\begin{itemize}
\item \textbf{VulDeeLocator}~\cite{VulDeeLocator}: 
a Bidirectional Recurrent Neural Network (BiRNN) based model by intermediate representation (IR) to locate vulnerable lines of code. It uses a slicing method similar to SySeVR.

\item \textbf{VulChecker}~\cite{vulchecker}:
A graph-based neural network  that utilizes Structure2Vec to identify subgraphs corresponding to real bugs. Nodes in the enriched program dependency graph (ePDG) are represented using LLVM~\cite{LLVM} IR.
\item \textbf{IVDetect}~\cite{IVDetect}: Subgraphs are selected by interpreting the classification results of Graph Convolution Network with feature-attention (FA-GCN) using an interpretable method GNNExplainer~\cite{GNNExplainer} to locate vulnerable nodes.   
\end{itemize}

For Linevul and Devign, we utilize the line of code and the node with the highest predicted probability, respectively as the results for bug localization. Based on Big-Vul and SARD, we conduct two datasets including bug localzation labels in our experiments. Across different localization methods, our model consistently outperforms others, including GNNExplainer which is used in IVDetect, achieving a 4.5\% improvement in \(IoU\). In contrast, line-level localization methods like Linevul may only provide isolated code line information, which makes it challenging to deliver a comprehensive analytical perspective. Our model generally performs better on Big-Vul, demonstrating at least an 11\% improvement in accuracy and a minimum of 2.18\% improvement in \(IoU\). 

Concurrency bug patterns are frequently linked to particular code structures, and our experiments demonstrate that the SubgraphX effectively captures these structures. Figure \ref{fig:location motivation} presents an example (data race) in which multiple nodes have been identified as containing concurrency bugs. Utilizing subgraph representations allows for a more intuitive visualization of bugs and their affected components, thereby enhancing the interpretability of detection results.

Concurrency bugs often arise from interactions across multiple nodes forming subgraphs, rather than isolated nodes.  By prioritizing the exploration of subgraphs likely to exhibit concurrency patterns, Convul enhances efficiency and reduces computational overhead. The approach enables precise and interpretable localization of vulnerable code regions by effectively capturing complex code dependencies in multi-threaded contexts, outperforming traditional methods. 

\section{Conclusion}
We propose Convul, a DL based approach for concurrency bug detection and localization that leverages CodeBERT, RGCN, and SubgraphX. On average, Convul demonstrates an significant improvement compared to alternative methods across diverse evaluation settings for concurrency bug detection and localization.
Convul transforms each concurrent program into a CCPG with a representation of the concurrency relationship. Our approach demonstrates the effectiveness of static program analysis combined with DL methods.
The limitation is that the algorithm for concurrency edges is only applicable to C/C++ multi-threaded programs. 

\section{DATA AVAILABILITY}
The data underlying this article will be shared on reasonable
request to the corresponding author.





\bibliographystyle{unsrt}
\bibliography{oup-authoring-template}

\end{document}